\let\texyear\year

\documentclass{ieeeaccess}
\usepackage{amsmath,amssymb,amsfonts}
\usepackage{algorithmic}
\usepackage{graphicx}
\usepackage{textcomp}
\usepackage[numbers, sort]{natbib}
\usepackage{bm}
\let\ieeeaccessyear\year
\let\year\texyear

\usepackage{tikz}
\usepackage{adjustbox}
\usepackage[edges]{forest}
\usepackage{xcolor}
\usepackage{ragged2e} 
\usepackage{varwidth}
\usetikzlibrary{trees, arrows.meta, positioning, patterns, calc, shapes.geometric}
\usepackage{graphicx}      
\let\year\ieeeaccessyear
\definecolor{accessblue}{cmyk}{1, 0.3, 0, 0.2}
\definecolor{greycolor}{cmyk}{0,0,0,.8}
\definecolor{colRetrieval}{RGB}{0, 85, 164}
\definecolor{bgRetrieval}{RGB}{235, 242, 250}

\definecolor{colGranularity}{RGB}{0, 128, 96}
\definecolor{bgGranularity}{RGB}{235, 248, 244}

\definecolor{colOrchestration}{RGB}{184, 60, 20}
\definecolor{bgOrchestration}{RGB}{252, 240, 235}

\definecolor{colRobustness}{RGB}{102, 51, 153}
\definecolor{bgRobustness}{RGB}{245, 240, 250}

\definecolor{colRoot}{RGB}{50, 50, 50}
\newcommand{\treeheader}[2]{%
    {\color{#1}\large\textbf{#2}}\par\smallskip \\%
}
\newcommand{\treeitemtitle}[1]{\textbf{#1}}
\tikzset{
    queryPoint/.style={circle, fill=blue!80!white, draw=black, thin, minimum size=10pt, inner sep=0pt},
    docPoint/.style={rectangle, fill=red!80!white, draw=black, thin, minimum size=10pt, inner sep=0pt},
    queryCenter/.style={circle, draw=blue!40, dashed, thick, fill=blue!5, minimum size=28pt, inner sep=0pt, font=\sffamily\bfseries, text=blue!50},
    docCenter/.style={circle, draw=red!40, dashed, thick, fill=red!5, minimum size=28pt, inner sep=0pt, font=\sffamily\bfseries, text=red!50},
    annotationText/.style={font=\sffamily\scriptsize, align=center, color=black!80},
    captionText/.style={font=\sffamily, align=justify, text width=9cm, color=black!90}
}
\makeatletter
\AtBeginDocument{\DeclareMathVersion{bold}
\SetSymbolFont{operators}{bold}{T1}{times}{b}{n}
\SetSymbolFont{NewLetters}{bold}{T1}{times}{b}{it}
\SetMathAlphabet{\mathrm}{bold}{T1}{times}{b}{n}
\SetMathAlphabet{\mathit}{bold}{T1}{times}{b}{it}
\SetMathAlphabet{\mathbf}{bold}{T1}{times}{b}{n}
\SetMathAlphabet{\mathtt}{bold}{OT1}{pcr}{b}{n}
\SetSymbolFont{symbols}{bold}{OMS}{cmsy}{b}{n}
\renewcommand\boldmath{\@nomath\boldmath\mathversion{bold}}}
\makeatother

\def\BibTeX{{\rm B\kern-.05em{\sc i\kern-.025em b}\kern-.08em
    T\kern-.1667em\lower.7ex\hbox{E}\kern-.125emX}}

\begin{document}

\title{Taxonomy of the Retrieval System Framework: Pitfalls and Paradigms}
\author{\uppercase{Deep Shah}\authorrefmark{1},
\uppercase{Sanket Badhe}\authorrefmark{2}, and \uppercase{Nehal Kathrotia}\authorrefmark{3}}

\address[1]{Google LLC, CA 94043 USA (e-mail: shahdeep@google.com)}
\address[2]{Google LLC, CA 94043 USA (e-mail: sanketbadhe@google.com)}
\address[3]{Google LLC, CA 94043 USA (e-mail: nehalk@google.com)}

\markboth
{Shah \headeretal: Taxonomy of the Retrieval System Framework:
Pitfalls and Paradigms}
{Shah \headeretal: Taxonomy of the Retrieval System Framework:
Pitfalls and Paradigms}

\corresp{Corresponding author: Deep Shah (e-mail: shahdeep@google.com).}

\begin{abstract}
Designing an embedding retrieval system requires navigating a complex design space of conflicting trade-offs between efficiency and effectiveness. This work structures these decisions as a vertical traversal of the system design stack. We begin with the Representation Layer by examining how loss functions and architectures, specifically Bi-encoders and Cross-encoders, define semantic relevance and geometric projection. Next, we analyze the Granularity Layer and evaluate how segmentation strategies like Atomic and Hierarchical chunking mitigate information bottlenecks in long-context documents. Moving to the Orchestration Layer, we discuss methods that transcend the single-vector paradigm, including hierarchical retrieval, agentic decomposition, and multi-stage reranking pipelines to resolve capacity limitations. Finally, we address the Robustness Layer by identifying architectural mitigations for domain generalization failures, lexical blind spots, and the silent degradation of retrieval quality due to temporal drift. By categorizing these limitations and design choices, we provide a comprehensive framework for practitioners to optimize the efficiency-effectiveness frontier in modern neural search systems.
\end{abstract}


\titlepgskip=-21pt

\maketitle

\section{Introduction} \label{sec:introduction}

The paradigm of Information Retrieval (IR) has undergone a fundamental transformation, shifting from the rigid exact match mechanics of inverted indices to semantic matching driven by high dimensional vector spaces. This evolution, accelerated by the advent of Pre-trained Language Models (PLMs) and Large Language Models (LLMs), has redefined the core objective of retrieval from identifying documents containing specific query terms to identifying content that satisfies complex and implicit user intents \cite{zhu2025large}. Generative models often face hallucinations, precision issues, and staleness due to training date cut-offs \cite{cheng2024dated} and an over-reliance on parametric knowledge \cite{yu2024revealing}. Retrieval-Augmented Generation (RAG) \cite{lewis2020retrieval} has shown promising results in mitigating hallucinations \cite{li2025mitigating} and improving the reasoning capability \cite{liu2024much} of downstream applications.

Today, dense retrieval serves as the backbone for critical generative applications across sectors such as Finance \cite{Chen2025-ih, Iaroshev2024-tl}, Healthcare \cite{Westbrook2005, Tu2026}, Software \cite{unterkalmsteiner2016large, zhao2025understandingdesigndecisionsretrievalaugmented}, and Legal \cite{reuter2025reliableretrievalragsystems}. Given the widespread adoption of embedding-based retrieval systems across domains with varying complexity and requirements, practitioners must navigate a Pareto frontier of efficiency and effectiveness where optimizing one dimension often necessitates compromising another. For instance, while Bi-encoder architectures offer the scalability required for billion-scale indexing via Maximum Inner Product Search (MIPS) \cite{karpukhin2020dense}, they inherently suffer from a representation bottleneck. Compressing complex documents into a single vector inevitably results in information loss, making them structurally blind to fine-grained nuances that computationally expensive Cross-encoders can capture \cite{reimers2019sentence, nogueira2019passage}.

Beyond the architectural choice of the encoder, the efficacy of a retrieval system is heavily contingent on decisions made throughout the entire system stack. The granularity at which documents are segmented, whether via fixed windows, semantic boundaries \cite{verma2025s2}, or hierarchical trees \cite{sarthi2024raptor}, determines the semantic coherence of the retrieval unit. Similarly, the static nature of standard training paradigms often leaves models vulnerable to silent degradation caused by temporal drift, where the semantic alignment of queries and documents decays as language and facts evolve over time \cite{lazaridou2021mind}. Designing robust embedding-based retrieval systems requires more than training a proficient encoder model; it demands the careful design of distinct albeit interconnected components.

\begin{figure*}[!t]
  \centering
  \begin{adjustbox}{max width=\textwidth, max height=0.9\textheight, keepaspectratio}
    \begin{forest}
      for tree={
        grow=0,                 
        reversed,               
        forked edges,           
        draw,
        rounded corners=4pt,
        node options={inner sep=8pt, align=left},
        l sep=12mm,             
        s sep=6mm,              
        edge={draw=darkgray, line width=1pt, rounded corners=2pt},
        font=\sffamily,
      },
      root/.style={
        fill=colRoot,
        text=white,
        font=\Large\bfseries\sffamily,
        text width=3.5cm,
        align=center,
        anchor=west
      },
      layer/.style 2 args={
        fill=#1,
        draw=#1,
        text=white,
        font=\bfseries\large,
        align=center,
        anchor=west,
        edge path={
          \noexpand\path[\forestoption{edge}]
            (!u.parent anchor) -- ++(5mm,0) |- (.child anchor)\forestoption{edge label};
        },
        for children={
            contentnode={#1}{#2}
        }
      },
      contentnode/.style 2 args={
        fill=#2,
        draw=#1,
        line width=1.2pt,
        text=black,
        font=\small\rmfamily,
        anchor=west,
        child anchor=west,
      }
      [Retrieval\\ System\\Framework, root
        [Representation\\Layer, layer={colRetrieval}{bgRetrieval}
          [{\treeheader{colRetrieval}{Architecture}
            \treeitemtitle{Dual Encoders} \cite{karpukhin2020dense, reimers2019sentence}, 
            \treeitemtitle{Cross-Attention} \cite{devlin2019bert, nogueira2019passage}, 
            \treeitemtitle{Late Interaction} \cite{khattab2020colbert, santhanam2022colbertv2}}]
          [{\treeheader{colRetrieval}{Loss Function}
            \treeitemtitle{Contrastive} \cite{pmlr-v195-parulekar23a, hadsell2006dimensionality}, 
            \treeitemtitle{Triplet Margin} \cite{Schroff_2015, hoffer2018deepmetriclearningusing}, 
            \treeitemtitle{Listwise} \cite{cao2007learning, qin2010generalapproximation}}]
          [{\treeheader{colRetrieval}{Negative Sampling}
            \treeitemtitle{Static Hard Negative} \cite{nguyen2023passage}, 
            \treeitemtitle{Curriculum} \cite{xiong2020approximate, santosh2024cusines}, 
            \treeitemtitle{False Negative Filter} \cite{qu2021rocketqa, robinson2020contrastive}}]
        ]
        [Granularity\\Layer, layer={colGranularity}{bgGranularity}
          [{\treeheader{colGranularity}{Fixed Chunking}
            \treeitemtitle{Token Window} \cite{lewis2020retrieval, Safjan_2023}, 
            \treeitemtitle{Global Context} \cite{anthropicContextualRetrieval, Voyage_AI_2025}}]
          [{\treeheader{colGranularity}{Semantic Chunking}
            \treeitemtitle{Embedding Shifts} \cite{verma2025s2, kiss2025max}, 
            \treeitemtitle{Model Segmentation} \cite{tripathi2025vision, glasz2025can}}]
          [{\treeheader{colGranularity}{Atomic Chunking}
            \treeitemtitle{Atomic Facts} \cite{chen2024dense}}]
          [{\treeheader{colGranularity}{Hierarchical Chunking}
            \treeitemtitle{Recursive Summaries} \cite{sarthi2024raptor, wu2022learning}, 
            \treeitemtitle{Structural Trees} \cite{lu2025hichunk, hoai2025data}}]
        ]
        [Orchestration\\Layer, layer={colOrchestration}{bgOrchestration}
          [{\treeheader{colOrchestration}{Transcending the Embedding Bottleneck}
            \textit{Multi-vector:} \treeitemtitle{Multi-View} \cite{luan2021sparse, dhulipala2024muvera}, \treeitemtitle{Autoregressive} \cite{chen2025beyond} \\
            \textit{Fanout:} \treeitemtitle{Query Decomposition} \cite{jin2025searchr1trainingllmsreason, zhao2025parallelsearch}, \treeitemtitle{Confidence Loops} \cite{jiang2023activeretrievalaugmentedgeneration, long2025divermultistageapproachreasoningintensive} \\
            \textit{Agentic:} \treeitemtitle{Multi-hop Traversal} \cite{gupta2025llm} \\
            \textit{Hierarchical:} \treeitemtitle{LLM Guided} \cite{gupta2025llm}, \treeitemtitle{Recursive Summary} \cite{sarthi2024raptor}}]
          [{\treeheader{colOrchestration}{Reranking Pipeline}
            \textit{Pipeline:} \treeitemtitle{Cascade Filters} \cite{wang2011cascade, cambazoglu2010early}, \treeitemtitle{Distillation} \cite{choi2021improving, bhattacharya2023cupid}. \\
            \textit{Precision:} \treeitemtitle{Listwise Sort} \cite{pradeep2023rankzephyr, esfandiarpoor2025beyond}, \treeitemtitle{Reasoning CoT} \cite{weller2025rank1, guo2025deepseek}. \\
            \textit{Objectives:} \treeitemtitle{Diversity} \cite{carbonell1998use, yan2021diversification}, \treeitemtitle{Fairness} \cite{wu2021fairness}}]
        ]
        [Robustness\\Layer, layer={colRobustness}{bgRobustness}
          [{\treeheader{colRobustness}{Domain Generalization}
            \treeitemtitle{Zero-Shot Transfer} \cite{thakur2021beir}, 
            \treeitemtitle{Task Instructions} \cite{su2023one, wang2024improving}}]
          [{\treeheader{colRobustness}{Lexical Gap}
            \treeitemtitle{Learned Sparse} \cite{formal2021splade, chen2022salient}, 
            \treeitemtitle{Vector Limits} \cite{weller2025theoretical, tishby2000information}}]
          [{\treeheader{colRobustness}{Temporal Shift}
            \treeitemtitle{Distribution Decay} \cite{lazaridou2021mind, han-etal-2025-temporal}, 
            \treeitemtitle{Timestamp Injection} \cite{abdallah2025tempretriever, dhingra2022time}, 
            \treeitemtitle{Drift Compensation} \cite{goswami2025querydriftcompensationenabling}}]
        ]
      ]
    \end{forest}
  \end{adjustbox}
  \caption{Taxonomy of the Retrieval System Framework.}
  \label{fig:retrieval_framework}
\end{figure*}
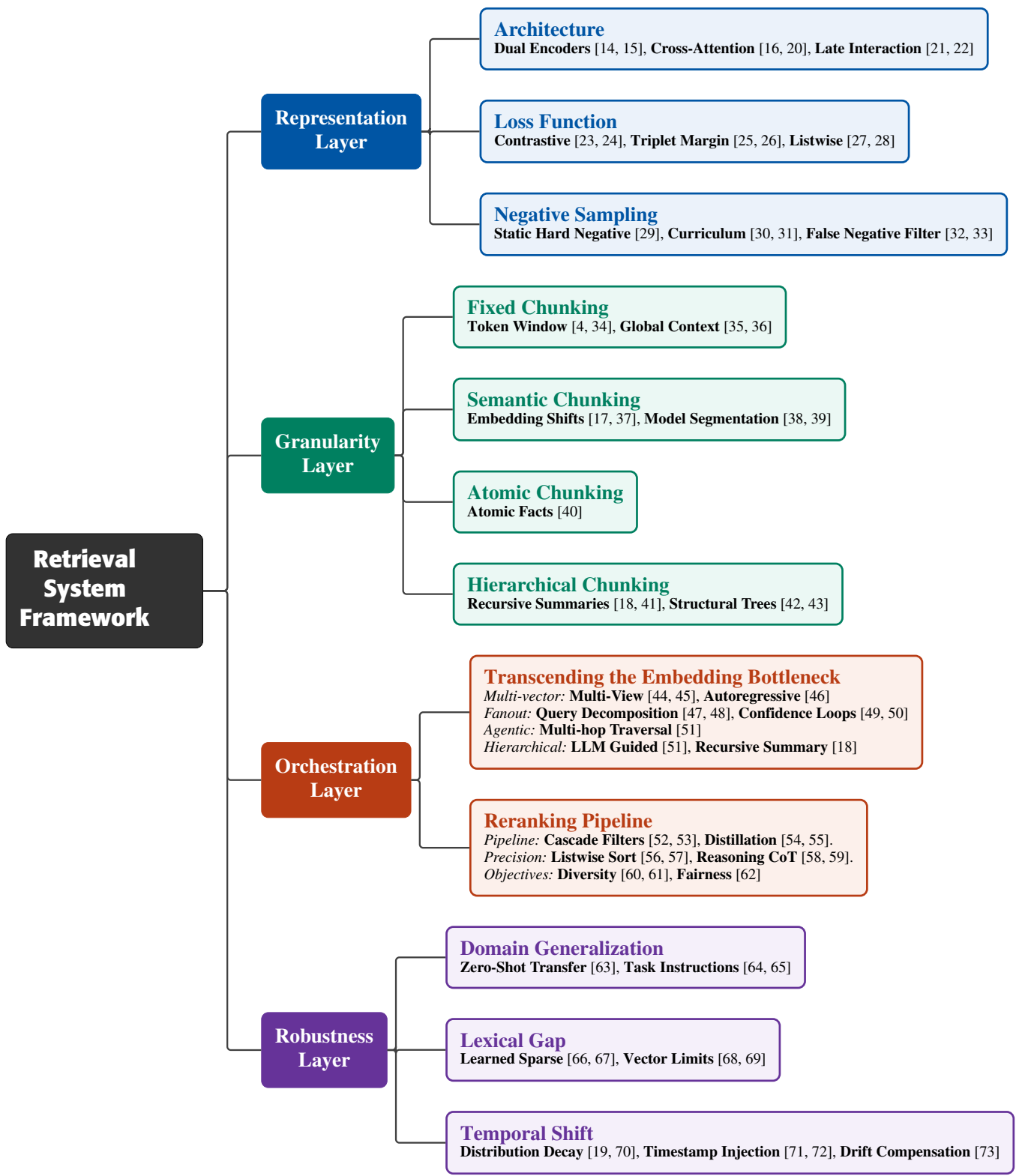

This article analyzes the components involved in the embedding-based retrieval system. We propose a framework (Fig. \ref{fig:retrieval_framework}) that categorizes design decisions into four distinct layers, analyzing how choices at one level constrain optimization at others:
\begin{enumerate}
    \item \textbf{The Representation Layer:} We examine the foundational trade-offs in loss functions and architectural topologies. We analyze the dichotomy between the efficiency of Bi-encoders and the expressivity of Cross-encoders, highlighting hybrid Late Interaction paradigms that attempt to bridge this gap.
    \item \textbf{The Granularity Layer:} We evaluate how document segmentation impacts the information bottleneck. We contrast atomic chunking strategies against hierarchical approaches \cite{sarthi2024raptor}, demonstrating how the semantic coherence of the retrieval unit impacts downstream generation.
    
    \item \textbf{The Orchestration Layer:} We discuss optimizations that transcend the single-vector paradigm, including query fanout, agentic decomposition, and multi-stage reranking pipelines. This section focuses on resolving capacity limitations through dynamic retrieval logic.
    
    \item \textbf{The Robustness Layer:} Finally, we address the critical challenges of domain generalization, lexical entity matching, and temporal validity. We identify architectural mitigations for degradation caused by temporal drift, ensuring systems remain reliable in evolving data environments.
\end{enumerate}

By structuring these limitations and design choices into a cohesive stack, this work provides a comprehensive guide for researchers and practitioners aiming to optimize the efficiency-effectiveness frontier in next-generation neural search systems. To maintain a focused analysis on the algorithmic optimization of the embedding-based retrieval subsystem, this review explicitly delimits its scope to text-based dense retrieval, excluding multimodal retrieval, knowledge graphs, and low-level index engineering.

\let\ieeeaccessyear\year
\let\year\texyear
\let\year\ieeeaccessyear

\subsection{Architecture}

\subsubsection{Bi-Encoders Architecture}
The asymmetric dual encoder architecture, often categorized within bi-encoder frameworks, employs two distinct encoders: a query encoder ($E_Q$) and a document encoder ($E_D$) to project inputs into a unified, low-dimensional dense semantic space \cite{karpukhin2020dense}. In Transformer-based implementations, these models typically derive a fixed-length representation via the \texttt{[CLS]} token or mean pooling operations. In contrast to symmetric Siamese networks \cite{bromley1993signature}, this architecture utilizes an asymmetric parameterization with unshared weights. This configuration enables $E_Q$ to specialize in short, keyword-centric syntax while $E_D$ adapts to the structural nuances of longer, expository text.

Relevance between query-document pairs is quantified through a dot product, a crucial choice that allows the integration of various vector-based indexing systems for scalable retrieval. This decoupling of encoders allows for the offline pre-computation and indexing of the document corpus, reducing online inference to a single forward pass of $E_Q$ followed by a Maximum Inner Product Search (MIPS). While this paradigm significantly minimizes retrieval latency, it inherently introduces a representation bottleneck. The requirement to compress complex semantic interactions into a single vector inevitably results in missing the fine-grained and detailed interaction between the query and the document.

\subsubsection{Cross-Encoder Architecture}
Unlike the Asymmetric Dual Encoder, which enforces a strict separation between query and document representations until the final scoring step, the Cross-Encoder architecture processes the query and document as a single, contiguous sequence. This architecture serves as the upper bound for retrieval effectiveness on the efficiency-effectiveness frontier.

The query and document are concatenated into a single input sequence, typically separated by special tokens (e.g., [CLS] Query [SEP] Document \cite{devlin2019bert}). This combined sequence is fed into a single Transformer encoder. Critically, this allows the self-attention mechanism to attend to every token in the query with respect to every token in the document across all transformer layers. This early interaction allows the model to capture deep semantic dependencies, negation logic \cite{petcu2025comprehensive}, and exact matching \cite{lu2025pathway} that are often lost when compressing text into a bottlenecked vector representation. It effectively resolves the representation bottleneck inherent to bi-encoders.

However, the superior expressivity of cross-encoders comes with a prohibitive computational cost that scales linearly with the corpus size. Because the document representation cannot be pre-computed and cached as it depends on the specific query tokens, scoring a query against a corpus of size $N$ requires $O(N)$ forward passes of the heavy transformer model. Due to this extreme latency, cross-encoders are computationally infeasible for first-stage retrieval in large-scale systems. Instead, they are typically reserved for high-precision tasks such as Reranking top candidates or as Teacher models to distill knowledge into more efficient student bi-encoders \cite{choi2021improving, bhattacharya2023cupid}.

\subsubsection{Hybrid Architectures}
\begin{figure}
    \centering
    \includegraphics[width=1.0\linewidth]{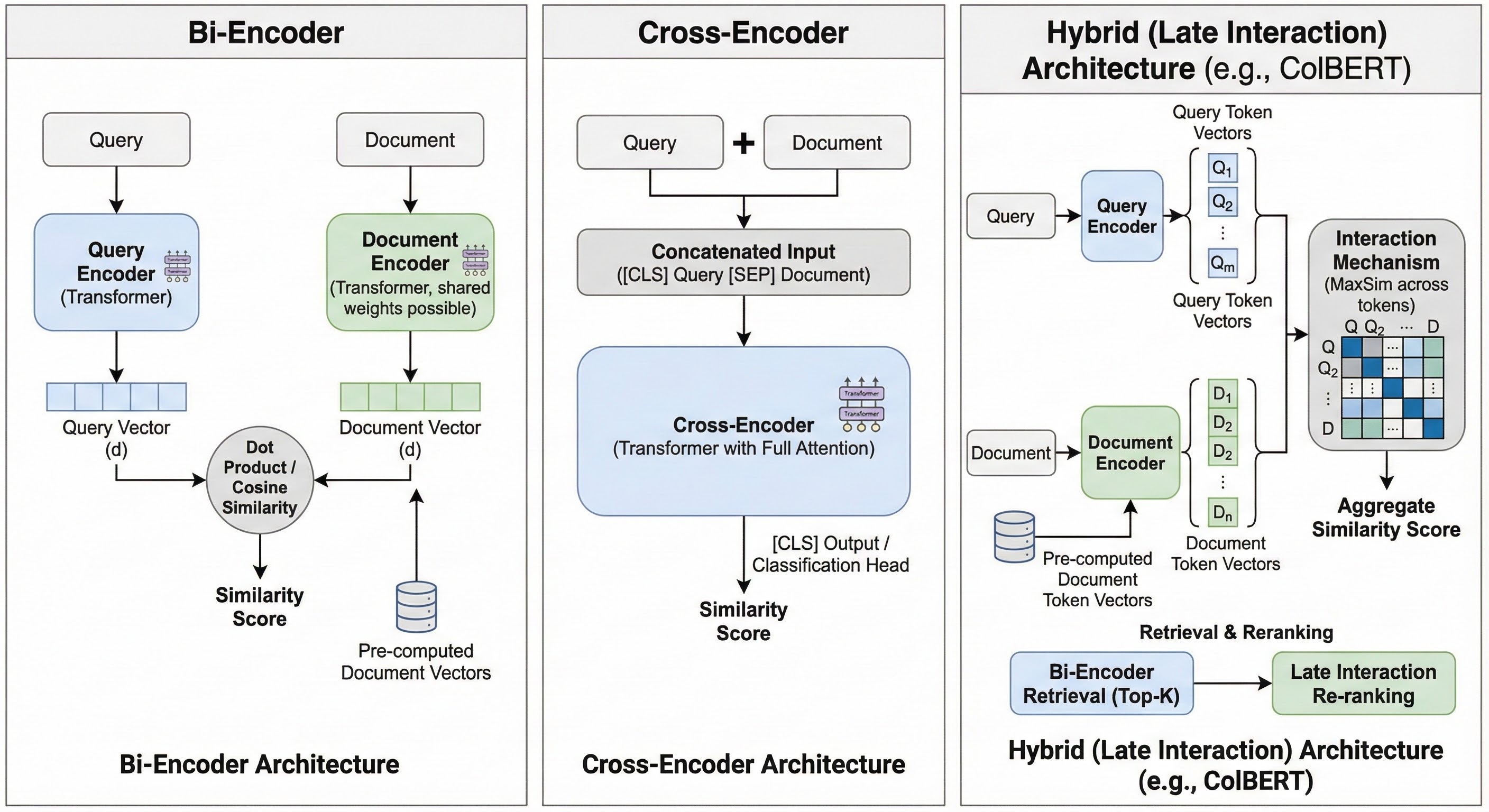}
    \caption{Comparison of different retrieval architectures: Bi-encoder vs Cross-encoder vs  action architecture}
    \label{fig:architecture_comparision}
\end{figure}

To bridge the substantial efficiency-effectiveness gap between bi-encoders and cross-encoders, hybrid architectures have emerged. As shown in Fig. \ref{fig:architecture_comparision}, these approaches \cite{humeau2019poly, khattab2020colbert} seek to retain the offline pre-computation capability of bi-encoders while introducing a mechanism for richer, non-linear late interaction between query and document representations to approximate the performance of cross-encoders.

One prominent approach is the Poly-encoder \cite{humeau2019poly}, which modifies the bi-encoder framework to learn a set of global features rather than a single vector. While the candidate label is still encoded into a single vector to allow for caching, the input context is projected into $m$ global vectors using learned attention codes. During inference, the candidate encoding acts as a query to attend over these $m$ context vectors, extracting a final context representation tailored to the specific candidate. This architecture performs a final layer of attention-based interaction, capturing more nuanced dependencies than a simple dot product while avoiding the prohibitive cost of full cross-attention at every layer.

Alternatively, ColBERT (Contextualized Late Interaction over BERT) \cite{khattab2020colbert} introduces a Late Interaction paradigm. Instead of compressing documents into a single vector or learning global codes, ColBERT encodes queries and documents into bags of token-level embeddings. Relevance is computed via a MaxSim operator where, for every query token, the model identifies the most similar document token (maximum similarity) and sums these scores. This retains the fine-grained, token-level matching capability of cross-encoders but ensures that document embeddings can still be pre-computed and stored offline.

Recent advancements, such as ColBERTv2 \cite{santhanam2022colbertv2}, further refine this approach by addressing the storage overhead of multi-vector representations. By employing residual compression, storing vectors as low-bit residuals from learned centroids, and utilizing denoised supervision strategies, these models achieve state-of-the-art retrieval quality with a storage footprint competitive with single-vector models.

\subsection{Loss Function}

\subsubsection{Contrastive Loss}
Standard contrastive loss \cite{hadsell2006dimensionality, pmlr-v195-parulekar23a} is a pairwise loss operating on two types of data points: positive pairs that should be similar and negative pairs that should be dissimilar. The objective function minimizes the distance between positives and maximizes the distance between negatives. Since the input consists strictly of positive and negative pairs, the embeddings of various positives for a query are pushed closer towards the query even after a reasonable margin of separation has been reached. This leads to cluster collapse, which is discussed later.

InfoNCE (Information Noise-Contrastive Estimation) \cite{pmlr-v195-parulekar23a} is a contrastive loss function used to learn high-quality data representations by maximizing a lower bound on the mutual information (MI) between different views of the same data. Computing InfoNCE for each sample is computationally infeasible during training; therefore, \textbf{In-Batch negatives} serve as approximations. This method leverages the batch structure itself for efficiency. For every query in a training batch of size $B$, the specifically paired document is treated as the positive sample, while the other $B - 1$ documents paired with other queries in the same batch serve as implicit negative samples.

\textit{\textbf{In-Batch negatives}} are computationally efficient as they allow for massive effective negative sampling without the memory overhead of encoding specific negative samples for every query. This approach creates a scaling law where performance generally improves with batch size, as larger batches provide a more diverse and representative set of "random" negatives that better approximate the true distribution of the corpus. In large batches or narrow domains, the batch is statistically likely to encounter false negatives. This impacts model alignment \cite{zhou2022debiased}. Various approaches to mitigate false negatives during negative sampling are discussed in Section \ref{negative_sampling}.

\subsubsection{Triplet Loss}
In contrast to contrastive loss, triplet loss accepts three inputs: a query, a positive document, and a negative document \cite{Schroff_2015, hoffer2018deepmetriclearningusing}. Similar to contrastive loss, the objective minimizes the distance between similar (query, document) pairs and maximizes the distance between dissimilar pairs. The objective function evaluates the distance between positive and negative documents against the query and ensures that the distance between the query and the positive document is smaller than the distance between the query and the negative document by at least a specified margin.

The crucial difference lies in the gradient computation. Once the triplet satisfies the margin constraint, the loss becomes zero. Consequently, the model ignores easy examples and preserves higher within-class diversity. Triplet loss prevents the model from the cluster collapse phenomenon discussed previously \cite{zeng2025comparing}.

\subsubsection{Listwise Loss}
Pairwise and triplet losses focus on local comparisons, which can lead to suboptimal global rankings. Listwise optimization (e.g., ListNet, ApproxNDCG) treats the entire list of retrieved documents as a single training instance \cite{cao2007learning, qin2010generalapproximation}. The objective maximizes the probability of the ideal permutation of documents given the query. This aligns the training objective more closely with evaluation metrics like NDCG@10. However, listwise ranking losses present several practical and research challenges, primarily related to training data generation, computational scaling, handling of noisy or incomplete data, and the complexity of integrating them effectively into existing machine learning systems. Using LLMs to generate training data for listwise loss is an active research direction \cite{esfandiarpoor2025beyond}.

\subsection{Negative Sampling} \label{negative_sampling}
The selection of a negative sampling strategy is often more critical to training a robust retriever than hyperparameters such as learning rate or model depth. A representation model learns to distinguish only what it is explicitly forced to differentiate. The literature identifies a taxonomy of negative hardness that dictates the learning curriculum. Practitioners often begin with less precise, computationally efficient negative sampling before progressing to more precise, computationally expensive strategies \cite{yang2021unsupervised, santosh2024cusines}.

A primary limitation of representation learning is cluster collapse, a phenomenon where learned representations collapse into dense groups rather than spreading across the embedding space, losing valuable discriminative information for downstream tasks. Another challenge is dimensional collapse \cite{jing2021understanding}, where embedding spaces collapse along certain dimensions rather than leveraging the entire manifold. Negative sampling strategies, particularly hard negatives, mitigate such collapses by forcing the model to retain such discriminative information \cite{yang2024does, xu2022negative, huang2023model}.

 Naive mining of hard negative samples frequently selects false negatives \cite{qu2021rocketqa}. \textbf{Denoising Hard Negatives} strategies propose filtering potential false negatives by identifying those with high relevance scores to encourage better representation learning \cite{chen2021incremental}. Powerful cross-encoder models or LLMs are employed to identify these false negatives \cite{qu2021rocketqa, lee2024gecko}. Detecting and introducing robustness against false negatives has demonstrated improvements \cite{robinson2020contrastive, chen2021incremental}.

\subsubsection{Random Negatives}
Random negatives are samples drawn arbitrarily from the corpus \cite{pan2008one, rendle2012bpr}. They are easy to distinguish and drive the initial phase of learning by teaching the model coarse-grained topic separation. However, relying solely on random negatives results in a model that fails to grasp nuance as it lacks exposure to hard negatives.

\subsubsection{Static Hard Negatives}
Static hard negatives involve using lexical retrieval systems like BM25 to identify hard negatives \cite{nguyen2023passage}. These are documents that share significant keyword overlap with the query but are not the ground-truth answer. Training against these forces the model to look beyond simple lexical matching and encode semantic intent. However, because they are static, the model can eventually memorize the specific biases of BM25, limiting further improvement.

\subsubsection{Dynamic Hard Negatives}
Dynamic hard negatives, such as the ANCE (Approximate Nearest Neighbor Negative Contrastive Estimation) \cite{xiong2020approximate} algorithm, introduce a dynamic curriculum that samples negatives using Approximate Nearest Neighbor (ANN) search. As the model trains, it utilizes its current state to retrieve negatives from the corpus. In early epochs, the model retrieves "semi-hard" negatives. As it improves, it retrieves increasingly difficult negatives that are semantically very close to the query but incorrect. This creates a natural, self-paced curriculum. Since rebuilding the index at each iteration is impractical and expensive, ANCE solves this by updating the index asynchronously, allowing the training loop to sample from a constantly refreshing distribution of difficulty.

\section{Document Chunking}
In Retrieval-Augmented Generation (RAG) systems, segmenting documents into smaller units, termed \textbf{chunks}, critically influences the quality of both retrieval and generation tasks \cite{chen2024dense, wadhwa2024rags, yu2024chain, pmlr-v202-shi23a}. Documents are rarely structured with retrieval optimization in mind, presenting a fundamental challenge due to their inherently low signal-to-noise ratio relative to a specific user query. This challenge is amplified in long-context documents. While such documents provide comprehensive coverage, typically only a fraction of the content is relevant for a given query. This disconnect introduces distinct failure modes in both retrieval and generation.

In the retrieval phase, projecting a multi-thematic document into a single dense embedding creates a representation bottleneck where granular details are diluted by dominant themes, significantly degrading recall for specific facts. In the generation phase, ingesting full-length documents remains suboptimal despite the expanded context windows of modern LLMs. Empirical research demonstrates the "Lost-in-the-Middle" phenomenon, where models exhibit a U-shaped performance curve and fail to extract relevant information situated away from context boundaries \cite{liu2024lost}. Furthermore, processing longer contexts incurs additional latency and token costs \cite{shekhar2024towards}. Chunking has proven effective for RAG applications on open-domain QA datasets \cite{lewis2020retrieval} and complex finance datasets containing 10-Ks and earnings reports \cite{islam2023financebench, yepes2024financial}.

To address these limitations, various chunking strategies have been proposed. These are broadly classified into Fixed, Semantic, Atomic, and Hierarchical approaches. Practitioners should view chunking not as a static preprocessing step but as a query-dependent variable. Documents are often indexed simultaneously at sentence, paragraph, and section levels to accommodate variance in query specificity, ranging from broad thematic searches to needle-in-a-haystack fact retrieval.

\subsection{Fixed Chunking}
Fixed chunking partitions input documents into uniform segments based on token count or page-level boundaries. \citet{lewis2020retrieval} first proposed RAG with fixed chunking of Wikipedia articles. A primary limitation of fixed chunking is the potential fracture of sentences and paragraphs across boundaries. The Sliding Window strategy mitigates this by retaining a subset of tokens from the previous chunk in the subsequent chunk \cite{Safjan_2023}. This ensures that sentences split across boundaries remain accessible in at least one retrieved segment.

When documents are divided into chunks, the overall document theme and surrounding context risk being lost, leading to suboptimal embedding representations. Generating and utilizing global document context prevents this loss \cite{Voyage_AI_2025}. To provide global context, metadata such as document summaries or titles are attached to each chunk. While helpful, global context risks causing embedding collapse if it dominates the chunk-specific content. Contextual Retrieval addresses this by prepending chunk-specific explanatory context to each segment before embedding \cite{anthropicContextualRetrieval}.

\subsection{Semantic Chunking}
Semantic chunking partitions documents based on natural breakpoints rather than arbitrary token counts. Breakpoints can be structured or unstructured. Structured breakpoints utilize natural demarcations such as code blocks, paragraphs, or section headers. Unstructured breakpoint methods compute semantic similarity between consecutive sentences and introduce a split if the similarity drops below a defined threshold \cite{kiss2025max, qu2025semantic}. \citet{verma2025s2} proposed merging spatial relationships and semantic similarity by creating a weighted graph followed by clustering.

Recent research has introduced multimodal LLM-guided chunking \cite{tripathi2025vision, glasz2025can}. This approach demonstrates expert human-level performance on documents with complex structures, content diversity, and visual elements \cite{tripathi2025vision}, as well as code summary generation tasks \cite{glasz2025can}. However, comparative studies indicate that semantic chunking does not consistently outperform fixed chunking on standard corpora, including long-document datasets like FinanceBench, as natural language transitions are often gradual rather than abrupt \cite{qu2025semantic}. Semantic chunking showed statistically significant gains primarily on datasets synthetically generated by stitching unrelated documents together. Consequently, the selection between semantic and fixed chunking should depend on the corpus type; for example, semantic chunking may offer advantages for corpora like news digests or meeting notes where distinct topic shifts are prevalent.

\begin{figure}
    \centering
    \includegraphics[width=1.0\linewidth]{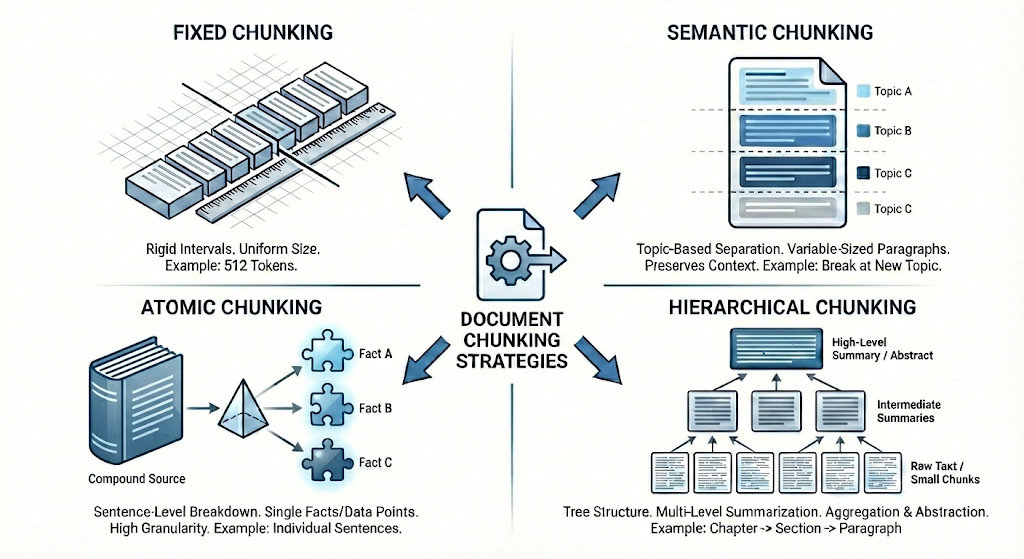}
    \caption{Taxonomy of document segmentation strategies categorized by granularity and context preservation. The classification contrasts rigid boundary methods (Fixed Chunking) with semantic-aware techniques (Atomic, Semantic, and Hierarchical Chunking) designed to mitigate information loss and maintain retrieval coherence.}
    \label{fig:hierarchical_chunking}
\end{figure}

\subsection{Atomic Chunking}
To address the limitations of fixed and semantic segmentation, Atomic Chunking extracts atomic, self-contained, and minimal facts that cannot be further split \cite{chen2024dense}. This strategy adheres to three key principles: distinct meaning, minimalism, and self-contained interpretation. The minimalist principle ensures that no two facts coexist within the same chunk; for instance, a compound sentence is split into independent segments to capture distinct attributes (e.g., location and height). The self-contained principle requires rewriting text to include necessary context, such as resolving pronouns to specific entities. These principles ensure that embeddings are not forced to represent overly broad chunks, thereby avoiding information bottlenecks while retaining context often missing in sentence-level chunking.

Atomic chunking has demonstrated consistent performance gains over passage and sentence chunking across various Question Answering datasets, including NQ \cite{kwiatkowski2019natural}, TQA \cite{joshi2017triviaqa}, SQuAD \cite{rajpurkar2016squad}, and EQ \cite{sciavolino2021simple} \cite{chen2024dense}. Beyond standard benchmarks, this granularity offers advantages in cross-task generalization. While performance gains are moderate on in-domain data, atomic chunking significantly outperforms passage retrieval on out-of-domain tasks and long-tail entity retrieval.

\subsection{Hierarchical Chunking}
Standard fixed-size or semantic chunking methods often fragment global context by processing text segments as isolated units. This limitation leads to significant performance degradation on tasks requiring holistic document understanding or when relevant evidence is distributed across distant sections \cite{lu2025hichunk, sarthi2024raptor}. Hierarchical chunking addresses this by constructing multi-level representations that preserve both granular details and high-level themes, allowing retrieval systems to access information at the level of abstraction most appropriate for a given query \cite{hoai2025data, sarthi2024raptor}. This methodology aligns with cognitive science models of representation learning, which suggest that grouping atomic units into larger chunks based on statistical dependence is essential for processing complex sequential data \cite{wu2022learning}.

Methodologies for constructing these hierarchies generally fall into two categories: recursive processing and structure-based segmentation. Recursive approaches utilize soft clustering algorithms like Gaussian Mixture Models to group embedding vectors, followed by LLM-based summarization to build a semantic tree from the bottom up \cite{sarthi2024raptor}. Conversely, structure-based frameworks leverage explicit document metadata or fine-tuned LLMs to predict logical segmentation points, creating hierarchies that reflect the inherent organization of the document \cite{lu2025hichunk, hoai2025data}. While hierarchical indexing incurs higher computational costs during construction, it demonstrates superior robustness in evidence-dense benchmarks, such as complex legal retrieval or thematic question answering, where it significantly outperforms flat baselines by reducing context fragmentation \cite{lu2025hichunk, hoai2025data, sarthi2024raptor}.

\section{Transcending the Embedding Bottleneck}
\subsection{Multi-View Document Representation}
Theoretical analyses suggest intrinsic limitations in the capacity of single-vector bi-encoders to represent complex documents. \citet{weller2025theoretical} identified sign-rank limitations where a fixed embedding dimension $d$ mathematically restricts the number of top-$k$ document combinations a model can distinguish. Similarly, \citet{luan2021sparse} argue that general dimensionality reduction theory prevents any single encoder from guaranteeing low inner product distortion without significant growth in encoding size. These geometric constraints become particularly acute in long documents where fixed-length dense encodings struggle to preserve the fine-grained ranking distinctions captured by high-dimensional sparse models.

\subsubsection{Fixed-Cardinality Encodings}
Late interaction architectures like ColBERT \cite{khattab2020colbert} and Poly-encoders \cite{humeau2019poly} offer high fidelity but have historically faced deployment challenges due to incompatibility with standard indexing systems. These models typically require specialized indices, such as PLAID \cite{santhanam2022plaid}, rather than standard MIPS-based solutions \cite{douze2025faiss}.

To circumvent these retrieval bottlenecks and enable the use of standard Approximate Nearest Neighbor (ANN) or Maximum Inner Product Search (MIPS), researchers proposed fixed-cardinality encodings. \citet{luan2021sparse} introduced ME-BERT (Multi-Vector Encoding from BERT), which utilizes the first $m$ token outputs of a deep transformer. Critically, ME-BERT defines relevance as the maximum dot product between the query vector $q$ and the set of document vectors $\{d_1, ..., d_m\}$:

\begin{equation}
s(q, d) = \max_{j=1...m} \langle q, d_j \rangle
\end{equation}

This specific scoring function allows the search to be implemented efficiently via standard ANN by adding $m$ entries per document to the index, eliminating the need for complex custom retrieval logic. Addressing the potential redundancy in naively selecting $m$ tokens, \citet{zhang2022multi} proposed the Multi-View Representation (MVR) framework. MVR inserts explicit viewer tokens ($[\text{VIE}_i]$) optimized via a global-local loss to ensure the views capture distinct semantic aspects while maintaining an ANN-compatible architecture. Overall, practitioners must weigh this expressivity against significant storage overhead, as storing $m$ vectors per document linearly scales index memory requirements.

\subsubsection{Retrieval Optimization for Late-Interaction Models}
While ME-BERT and MVR are designed for native MIPS compatibility, other high-performance late interaction models rely on the Chamfer Similarity operation. This operation is typically a sum of maximum similarities which does not naturally decompose into a simple MIPS problem.

To bridge this gap, \citet{dhulipala2024muvera} introduced MUVERA (Multi-Vector Retrieval Algorithm). Unlike previous optimizations restricted to specific models, MUVERA reduces the multi-vector similarity search of these complex late-interaction architectures back to single-vector search by generating Fixed Dimensional Encodings (FDEs). FDEs are constructed such that their dot product mathematically approximates the true multi-vector Chamfer similarity:

\begin{equation}
\langle \phi(q), \psi(d) \rangle \approx S_{\text{Chamfer}}(q, d)
\end{equation}

By providing a theoretical $\epsilon$-approximation of this similarity, MUVERA allows architectures like ColBERT \cite{khattab2020colbert} and its successors \cite{gao2021coil, lin2023fine, santhanam2022colbertv2, yao2021filip}, which previously required specialized retrieval mechanics, to utilize highly optimized off-the-shelf MIPS solvers. This makes them practical for large-scale retrieval systems.

\begin{figure}[htbp]
\centering
\begin{adjustbox}{width=\columnwidth}
\begin{tikzpicture}[scale=0.8]

    \draw[->, thick, color=black!60] (0,0) -- (8.5,0);
    \draw[->, thick, color=black!60] (0,0) -- (0,8.5);

    \coordinate (Q_center) at (2.5, 6.5);
    \coordinate (D_center) at (7.0, 2.0);

    \coordinate (q_connector) at (3.7, 5.3); 
    \coordinate (d_connector) at (5.8, 3.2); 

    \draw[dashed, very thick, black!70, shorten <=2pt, shorten >=2pt] (q_connector) -- (d_connector) 
        node[midway, above right, align=left, font=\sffamily\scriptsize\bfseries, 
             xshift=-4pt, yshift=0pt, color=black!80,
             fill=white, inner sep=2pt, fill opacity=0.9, text opacity=1] 
             {Reduced\\Semantic Gap};

    \node[queryCenter] at (Q_center) {Q};
    
    \node[queryPoint] at ($(Q_center) + (-0.9, 0.45)$) {};
    \node[queryPoint] at ($(Q_center) + (0.0, 1.2)$) {};
    \node[queryPoint] at ($(Q_center) + (-0.75, -0.75)$) {};
    \node[queryPoint] at ($(Q_center) + (1.05, 0.15)$) {};
    \node[queryPoint] (q_final) at (q_connector) {};

    \node[docCenter] at (D_center) {D};
    
    \node[docPoint] at ($(D_center) + (0.9, 0.75)$) {};
    \node[docPoint] at ($(D_center) + (0.45, -0.9)$) {};
    \node[docPoint] at ($(D_center) + (-0.9, -0.6)$) {};
    \node[docPoint] at ($(D_center) + (-0.15, 1.2)$) {};
    \node[docPoint] (d_final) at (d_connector) {};

    \node[annotationText, above right=1.5cm of Q_center, anchor=south] (q_note) {Query Fanout / Multi-Query Representation};
    \draw[->, thin, gray, shorten >=3pt] (q_note.south) to[out=270, in=110] (Q_center);

    \node[annotationText, below right=1.5cm of D_center, anchor=north] (d_note) {Document Multi-Vector Representations};
    \draw[->, thin, gray, shorten >=3pt] (d_note.north) to[out=90, in=290] (D_center);

\end{tikzpicture}
\end{adjustbox}
\caption{{Multi-view representations transcend the single-vector information bottleneck by expanding the embedding space to bridge the semantic gap.}}
\end{figure}
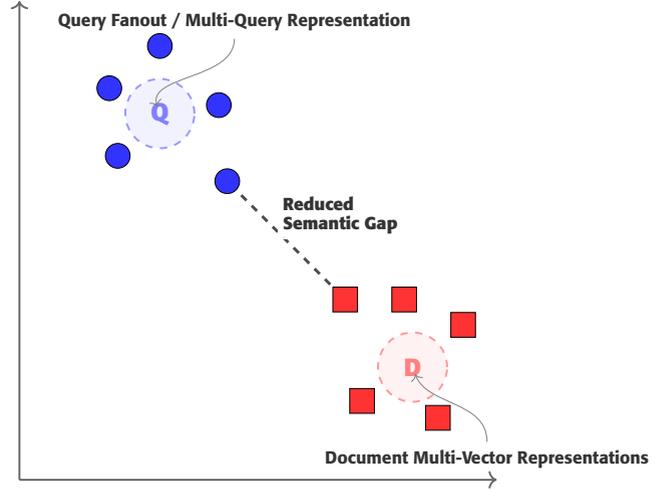

\subsection{Multi-Query Representation}
Standard dense retrieval methodologies predominantly operate under the assumption that the semantic relationship between a query $q$ and its relevant documents $\mathcal{D}^+$ can be modeled as a unimodal distribution within a shared vector space. In this paradigm, a bi-encoder is trained to minimize the distance between a single query embedding and the embeddings of ground-truth documents \cite{chen2025beyond}. However, complex information needs, particularly those arising from ambiguity or requiring list-based answers, often imply a multimodal target distribution where valid documents occupy disparate regions of the embedding space \cite{amouyal2022qampari, min2020ambigqa}.

When the set of relevant documents exhibits high pairwise distance, a single-vector query representation is mathematically forced to converge on the geometric centroid of the target clusters to minimize global loss. This averaging effect results in a query representation that, while central to the distribution, remains insufficiently close to any specific document to trigger an accurate retrieval match \cite{chen2025beyond}.

Architectural advancements in multi-view document representation are theoretically applicable here. Additionally, \citet{chen2025beyond} propose the \textbf{Autoregressive Multi-Embedding Retriever (AMER)}, a framework that reframes the encoding process as a sequence generation task within the continuous embedding space. Unlike traditional architectures that output a single static vector, AMER utilizes an autoregressive language model to generate a sequence of distinct query embeddings $\mathbf{Z} = \{\mathbf{z}_1, \dots, \mathbf{z}_k\}$.

The generation of each subsequent embedding $\mathbf{z}_{i+1}$ is conditioned on both the original query tokens and the history of previously generated embeddings:
\begin{equation}
    P(\mathbf{Z} | \mathbf{q}) = \prod_{i=1}^{k} P(\mathbf{z}_i | \mathbf{q}, \mathbf{z}_{<i})
\end{equation}
This mechanism effectively allows the model to traverse the latent space and capture multiple interpretations or aspects of the query without the computational overhead of independent inference passes \cite{chen2025beyond}. Empirical validation on multi-answer benchmarks (AmbigQA \cite{min2020ambigqa}, QAMPARI \cite{amouyal2022qampari}) confirms that this approach yields significant performance gains, particularly in subsets where the target documents exhibit high variance and are semantically distant from one another.

\subsection{Generative Query Decomposition and Agentic Retrieval}

While dense and multi-vector retrieval paradigms significantly improve recall for ambiguous queries, they face intrinsic limitations when addressing complex or multi-hop reasoning tasks where necessary evidence is stratified across disjoint documents. To resolve the structural mismatch between complex user intents and granular passages, retrieval architectures are shifting from direct semantic matching to \textit{Generative Query Decomposition} as a first step. Formally, this process maps a composite intent into a set of executable sub-queries (also known as query fanouts), often utilizing Large Language Models (LLMs) not merely as encoders but as reasoning engines.

\subsubsection{From Supervised Learning to Reinforcement Learning in Query Generation}
The lineage of query decomposition can be traced to early symbolic QA systems, such as IBM Watson's DeepQA, which utilized syntactic decomposition to resolve nested constraints \cite{ferrucci2010building}. Subsequent approaches relied heavily on supervised semantic parsing \cite{dong2016language, perez2020unsupervised} or dataset-specific schemas \cite{talmor2018web}. However, these methods suffered from domain brittleness as they failed to generalize outside their training distribution.

The advent of LLMs has introduced a paradigm shift towards unsupervised decomposition, leveraging zero-shot capabilities and Chain-of-Thought (CoT) prompting \cite{huang2023towards} to generalize synthesis logic across heterogeneous domains without task-specific architecture engineering \cite{Rackauckas_2024}.

Recent work has begun to optimize these decomposition policies via Reinforcement Learning (RL), moving beyond static prompting. By treating decomposition as a policy optimization problem, systems such as Search-R1 \cite{jin2025searchr1trainingllmsreason} and ParallelSearch \cite{zhao2025parallelsearch} utilize environmental feedback to align sub-query generation with downstream retrieval utility rather than immediate semantic similarity. This RL fine-tuning is critical for teaching the model to construct sub-tasks that yield structurally retrievable answers \cite{chan2024rq, zhong2025reasoning}.

\subsubsection{Independent Sub-Query Generation}
Strategies in this category address multifaceted queries where sub-tasks are effectively independent. Approaches such as RAG-Fusion \cite{Rackauckas_2024} and ParallelSearch \cite{zhao2025parallelsearch} utilize LLMs to generate diverse search vectors simultaneously and independently, aggregating results via Reciprocal Rank Fusion (RRF) to maximize coverage.

\subsubsection{Active Retrieval}
As a natural next step, for reasoning-intensive tasks where sub-queries exhibit sequential dependency (i.e., $q_{t+1}$ is conditional on the result $r_t$ of $q_t$), architectures have evolved into iterative frameworks. Paradigms such as Least-to-Most Prompting and Interleaving Retrieval with Chain-of-Thought (IRCoT) establish a dynamic feedback loop wherein retrieved facts explicitly ground subsequent reasoning steps \cite{trivedi2023interleavingretrievalchainofthoughtreasoning}. This interleaving significantly mitigates hallucinations by ensuring the generation trajectory is constrained by external evidence.

Recent methods have moved beyond retrieving at fixed intervals to confidence-driven mechanisms. These systems \cite{jiang2023activeretrievalaugmentedgeneration, long2025divermultistageapproachreasoningintensive} monitor token generation confidence, triggering retrieval only when the model detects a knowledge gap (e.g., low-probability tokens), effectively filtering noise and optimizing the utilization of the context window. A primary limitation of such confidence-based approaches is queries involving temporal shifts, where the model relies on parametric knowledge that is absent from recent updates.

\subsection{Hierarchical Retrieval}
Hierarchical retrieval frameworks have emerged as a robust solution by organizing document corpora into multi-level structures that capture both broad discourse and granular details \cite{gupta2025llm, sarthi2024raptor}. As discussed in the Hierarchical Chunking section, raw text chunks are clustered based on semantic similarity and summarized to form parent nodes. This process repeats recursively until a root node is formed, resulting in a hierarchy where higher levels represent thematic overviews and lower levels contain specific facts. This organization enables the system to capture distant dependencies within the corpus that are often lost in flat indexing, significantly improving performance on complex, multi-hop reasoning tasks.

Once the hierarchical index is established, determining the optimal traversal strategy is critical for retrieval efficiency. RAPTOR \cite{sarthi2024raptor} offers a collapsed tree approach that flattens the multi-layered structure into a single index for vector search. Empirical results suggest this offers greater flexibility in matching the correct level of granularity compared to rigid layer-by-layer traversal. RAPTOR also supports tree structure traversal where children are identified based on embedding similarity between the query and the node summary. LATTICE \cite{gupta2025llm} advances this paradigm by introducing agentic retrieval where, rather than relying on embedding similarity, Large Language Models (LLMs) actively navigate the structure using a greedy, best-first search. Efficient pruning is performed based on relevance scores. This agentic approach achieves a search complexity that is logarithmic in relation to the number of documents, making it computationally efficient even for large corpora where precision and reasoning are non-negotiable.

In terms of performance and cost, these hierarchical methods have demonstrated substantial improvements and state-of-the-art performance \cite{sarthi2024raptor, gupta2025llm} over traditional baselines on thematic question benchmarks (QuALITY \cite{pang2022quality}, NarrativeQA \cite{kovcisky2018narrativeqa}, QASPER \cite{dasigi2021dataset}) and reasoning benchmarks (BRIGHT \cite{su2024bright}).

The selection of a retrieval architecture, specifically balancing tree versus flat indexing and embedding-based versus LLM-guided traversal, must account for corpus size, latency constraints, and the required reasoning depth. However, a significant limitation of current hierarchical approaches is the reliance on static semantic trees. In scenarios involving dynamic corpora, pre-computed summaries of internal nodes do not update dynamically. This results in stale representations that misguide the traversal process.

\section{The Reranking Layer} \label{sec:reranking}
Efficient retrieval on billion-scale corpora often trades precision for recall. To bridge the gap between high-recall retrieval and high-precision user alignment, practitioners utilize the Reranking Layer. This layer typically operates on a reduced candidate set, which often ranges from top-50 to top-1000 documents. This reduction allows for the application of computationally intensive models that capture deep semantic interactions, logical reasoning, and multi-objective constraints that are infeasible during the initial indexing stage.

\subsection{Multi-stage Reranker Strategy}
The core design principle governing this strategy is the Cascade Ranking Model, first formalized by \citet{wang2011cascade}. This framework operationalizes the efficiency-effectiveness trade-off by subjecting progressively smaller candidate sets to a sequence of increasingly expensive ranking functions.

In a typical dense retrieval pipeline, the first stage reduces the search space from $N$ documents to a candidate set $C_1$ where $|C_1| \ll N$. The reranking layer then applies a scoring function $\phi$ such that the latency cost is justified by the precision gain. This multi-stage approach allows for coarse-grained first-stage retrieval through techniques such as quantization or Matryoshka representation learning \cite{kusupati2022matryoshka, pmlr-v119-chen20l}. However, the efficacy of this pipeline is limited by the recall ceiling of the first stage. No amount of sophisticated reranking can recover relevant documents that the initial bi-encoder failed to retrieve, which makes the first-stage retrieval quality critical.

\subsection{High-Precision Reranking Methodologies}
The methodologies for reranking have evolved from feature-based learning-to-rank algorithms \cite{burges2010ranknet} to deep interaction models and, more recently, Large Language Model (LLM) based reasoners.

\subsubsection{Deep Interaction Rerankers}
Deep interaction rankers, primarily Cross-encoders, represent the standard for high-precision semantic matching \cite{nogueira2019passage}. Unlike bi-encoders which score candidates based on vector similarity, Cross-encoders and their late-interaction variants \cite{khattab2020colbert, humeau2019poly} process the query and document as a simultaneous input sequence. This allows the self-attention mechanism to capture token-level interactions across the pair.

While highly effective, the computational cost of these models scales linearly with the number of candidates. Recent research seeks to mitigate this via distillation, where the reasoning capabilities of massive models are distilled into lighter interaction-based rankers \cite{choi2021improving, bhattacharya2023cupid}.

\subsubsection{Generative and LLM-based Rerankers}
The advent of LLMs has introduced a paradigm where reranking is treated as a generation or reasoning task rather than simple classification. \citet{sun2023chatgpt} and \citet{ma2023large} demonstrated that while LLMs may struggle with zero-shot retrieval over large corpora, they excel as rerankers, particularly for hard samples that require reasoning beyond semantic overlap.

Current approaches in this domain bifurcate into pointwise and listwise generation:
\begin{itemize}
    \item \textbf{Pointwise Reranking:} The LLM evaluates relevance for each query-document pair independently \cite{zhuang2024beyond, long2025precise}. However, this method lacks the context of the broader candidate set.
    \item \textbf{Listwise Reranking:} Models like RankZephyr \cite{pradeep2023rankzephyr} and other recent advancements \cite{esfandiarpoor2025beyond} ingest the entire list of candidates simultaneously. This allows the model to perform comparative reasoning by attending to the relative strength of evidence across documents.
\end{itemize}

The integration of reasoning capabilities into reranking is further exemplified by frameworks like \textit{DeepSeek-R1} \cite{guo2025deepseek} and \textit{Llm4rerank} \cite{gao2025llm4rerank}, which incentivize the model to generate intermediate reasoning steps before assigning a relevance score. \citet{weller2025rank1} recently analyzed the test-time compute scaling for reranking, suggesting that allocating more inference time for reasoning during the reranking phase yields diminishing returns after a certain threshold.

\subsection{Multi-Objective Optimization}
In production environments, pure semantic relevance is rarely the sole objective. The reranking layer serves as the critical orchestration point to inject business logic, diversity, and fairness constraints before the retrieval system responds.

Retrieving a list of semantically near-identical documents often degrades the quality of downstream applications, particularly for ambiguous or multi-intent queries. To address this, \citet{carbonell1998use} introduced Maximal Marginal Relevance (MMR), which reorders documents to maximize a linear combination of relevance and novelty. Modern iterations extend this to latent space diversification using distributed representations \cite{yan2021diversification}.

Beyond diversity, algorithmic fairness is a critical requirement. \citet{wu2021fairness} proposed the \textit{FairRec} framework, which employs decomposed adversarial learning and orthogonality regularization to remove sensitive attributes from the representation used for ranking. These mitigations ensure that the retrieval system does not reinforce historical biases present in the training data. Effective integration of these disparate objectives requires rigorous score calibration to satisfy the ultimate system objective.

\section{Structural and Distributional Robustness}
Having established the optimization strategies within the Representation, Granularity, and Orchestration layers, this analysis addresses the Robustness Layer. While dense retrieval (DR) architectures offer semantic flexibility superior to rigid term-matching, they are not a universal solution for information retrieval. The optimization of embedding spaces for semantic similarity introduces distinct fragilities, particularly when models are deployed outside their training distribution or when tasks necessitate precise lexical fidelity. This section analyzes the architectural pitfalls of dense bi-encoders by categorizing them into two primary failure modes: distributional overfitting, where models fail to generalize on new domains, and structural blindness, where geometric compression obliterates fine-grained entity distinctions and term matching. This framework distinguishes these static, intrinsic limitations from the degradation of retrieval quality caused by evolving data streams.

\subsection{The Domain Generalization Challenge}
The fundamental premise of embedding-based retrieval posits that queries and documents can be mapped into a shared and semantically rich vector space. However, despite this theoretical promise and widespread adoption, empirical analysis reveals that these mappings are frequently brittle and exhibit significant overfitting to the training distribution. The most comprehensive validation of this limitation is provided by the BEIR benchmark \cite{thakur2021beir}, which evaluates ten retrieval architectures across 18 heterogeneous datasets spanning diverse domains.

The evaluation uncovers a sharp contrast between in-domain and zero-shot performance. While dense architectures \cite{karpukhin2020dense, xiong2020approximate} achieve substantial gains over lexical baselines on in-domain data, they consistently underperform traditional probabilistic models like BM25 \cite{robertson2009probabilistic} in zero-shot scenarios. Specifically, on datasets having large domain-shifts, such as BioASQ \cite{krithara2023bioasq}, dense models struggle to generalize. This failure is attributed to the domain-specific shift in terminology and the inherent lexical bias of the training data \cite{thakur2021beir}. This phenomenon underscores a critical finding: in-domain superiority is a poor predictor of out-of-distribution (OOD) robustness. Therefore, a thorough understanding of the query distributions during inference is crucial when designing retrieval systems.

Mechanistically, this failure arises because the encoder compresses lexical understanding based on the statistical regularities of the training data. When resolving OOD data where common tokens may function as rare technical nomenclature, the model lacks the sparse safety net of exact term-matching that provides BM25 its robustness. Instead, the model attempts to project novel terms into the pre-learned geometric space, which often results in mapping them erroneously to unrelated concepts. As noted in \cite{thakur2021beir}, the model overfits the low-rank geometric structure of the training data. Consequently, the hyperplanes required to separate relevant documents in a novel domain cannot be constructed from the existing parameterization.

\subsection{The Lexical Gap and Entity Blindness}
Complementing the challenges of domain generalization is the structural limitation imposed by the fixed-dimensional bottleneck. The architecture of single-vector dense retrievers entails an aggressive dimensionality reduction. This process compresses variable-length textual attributes into a compact latent representation, which is typically 768 dimensions. This compression acts as a low-pass filter that preserves high-frequency semantic signals while often obfuscating the precise lexical features necessary for exact retrieval. This phenomenon, termed the lexical gap, manifests as an inability to resolve queries contingent on rare entities, serial numbers, or specific proper nouns.

\citet{sciavolino2021simple} formalized this deficiency through the EntityQuestions benchmark, demonstrating that dense retrievers exhibit severe popularity bias. While performance remains robust for head entities with high co-occurrence frequencies in the pre-training corpus, accuracy degrades quickly for tail entities. Empirical analysis reveals that dense models perform worse than sparse baselines on specific relation types \cite{sciavolino2021simple}. Mechanistically, the embedding model frequently captures the coarse-grained semantic type of the target, such as identifying that the answer requires a location, but fails to discriminate between entities with spatially proximal vector representations. This leads to the retrieval of semantically plausible but factually incorrect candidates.

This limitation is intrinsic to the geometric nature of dense retrieval. Unlike sparse architectures which leverage inverted indexes to enforce explicit term overlap, dense models must approximate term matching through vector proximity. As highlighted by analyses of the SPLADE \cite{formal2021splade} and SPAR \cite{chen2022salient} frameworks, standard bi-encoders lack the capacity to reliably perform these exact-match operations for salient phrases. Consequently, even if the encoder retains knowledge of a specific entity, the resolution of the vector space may be insufficient to separate the correct target from its nearest semantic neighbors. This results in hallucinations where the system retrieves documents that are topically relevant but lexically mismatched \cite{luan2021sparse}. Efforts to bridge this gap have led to the proposal of context-aware sparse models that retain the exact-match fidelity of inverted indexes while integrating the contextual understanding of transformer-based architectures \cite{dai2020context, formal2021splade}.

\subsection{Theoretical Representational Limits}
Beyond empirical failures, recent scholarship identifies hard theoretical boundaries that constrain the efficacy of the single-vector paradigm. \citet{weller2025theoretical} formally derive a connection between embedding dimensionality and the sign-rank of the query-relevance matrix, which challenges the assumption that scaling model parameters alone can resolve retrieval shortcomings. They demonstrate that for any fixed dimension $d$, there exist combinatorial sets of top-$k$ relevant documents that are mathematically impossible to represent via a single vector dot product. The validity of this bound is evidenced by the failure of state-of-the-art models on the LIMIT dataset, which is a syntactically simple benchmark designed to probe these combinatorial edges. This suggests a fundamental representational ceiling for the single-vector approach \cite{weller2025theoretical}.

The Information Bottleneck (IB) principle provides a complementary information-theoretic perspective on these failures. In a bi-encoder architecture, the model compresses a variable-length sequence into a fixed-size vector. The optimization maximizes predictive information regarding the relevant document while minimizing the mutual information $I$ \cite{tishby2000information}. While this compression efficiently encodes high-frequency semantic concepts, it also filters out low-frequency information such as IDs or rare technical terms that the model deems redundant. This process leads to Lexical Collapse.

Furthermore, the reliance on the dot product or unnormalized cosine similarity introduces topological anomalies. Unlike Euclidean distance, the dot product does not satisfy the properties of a metric space because it fails the triangle inequality and reflexivity. Research into high-dimensional geometry indicates that a significant proportion of the document collection may reside within the interior of the data convex hull. These hidden documents cannot physically be the nearest neighbor to any arbitrary query vector. This geometric occlusion renders them effectively irretrievable regardless of query relevance \cite{Krause_2023}. This phenomenon exacerbates recall failure, particularly for outlier documents that do not lie on the accessible surface of the embedding manifold.

\subsection{Architectural Mitigations}
Contemporary mitigation strategies have bifurcated along two primary axes. These include representational scaling, which leverages the reasoning capacities of Large Language Models (LLMs) to densify the embedding space, and topological diversification, which abandons the dense single-vector constraint in favor of sparse or multi-granular interactions.

The landscape of dense retrieval has been reshaped by the integration of generative LLMs. Analysis of the Massive Text Embedding Benchmark (MTEB) confirms that state-of-the-art performance is no longer the purview of specialized architectures but rather general-purpose LLMs adapted for representation learning \cite{muennighoff2023mteb}. Models such as NV-Embed-v2 \cite{lee2024nv} and E5-Mistral-7B \cite{wang2024improving} demonstrate that the primary driver of robustness is the sophistication of the pre-training recipe and diversity rather than simple parameter scaling. Specifically, \citet{wang2024improving} highlight that incorporating diverse synthetic training data allows smaller open-source encoders to internalize complex semantic dependencies. This data-centric approach effectively simulates a world-knowledge regularizer and improves zero-shot generalization significantly more than traditional contrastive scaling.

\subsubsection{Instruction-Tuned and Dynamic Representations}
A significant advancement in mitigating representational rigidity is the development of dynamic embedding architectures. Unlike traditional bi-encoders that map a document to a static point in vector space regardless of context, instruction-tuned models such as INSTRUCTOR \cite{su2023one} and E5-Mistral \cite{wang2024improving} introduce a conditioning mechanism. These systems accept the target text and a natural language task description. This allows the model to modulate its internal attention heads dynamically.

Mechanistically, the instruction acts as a control vector that effectively rotates and projects the resulting embedding within the high-dimensional latent manifold. By attending to the instruction tokens, the encoder emphasizes specific semantic subspaces relative to the user intent while suppressing orthogonal features. This capacity for adaptive projection addresses the limitations of monolithic representational embeddings. It allows a single set of weights to simulate the behavior of multiple task-specific experts. As noted by \citet{wang2024improving}, this versatility is due to model scale. The underlying Large Language Model possesses the sufficient parametric capacity to disentangle these conflicting semantic views, which is a feat unattainable by smaller BERT-based architectures.

\subsubsection{Sparse and Multi-Vector Architectures}
To circumvent the theoretical information bottleneck inherent in compressing entire documents into a monolithic vector, recent architectures have abandoned the single-vector constraint in favor of either multi-faceted or sparse representations. These approaches seek to reclaim the fine-grained lexical resolution lost during dense pooling.

One prominent direction involves Learned Sparse Retrieval, which was proposed by the SPLADE framework \cite{formal2021splade}. Rather than compressing semantics into a dense latent space, the model projects embeddings into a high-dimensional sparse vocabulary of approximately 30,000 dimensions. The model learns to perform implicit expansion by assigning non-zero weights to relevant terms that do not explicitly appear in the text \cite{formal2021splade}. This architecture effectively bridges the lexical gap. It retains the exact-match fidelity of an inverted index while integrating the contextual capabilities of transformer-based language understanding.

Alternatively, Late Interaction architectures address the bottleneck by preserving the token-level embeddings of the encoder. ColBERT \cite{khattab2020colbert, santhanam2022colbertv2} exemplifies this paradigm. It defers the query-document interaction until runtime via an efficient MaxSim operator. By maintaining a multi-vector representation for each document, the system preserves fine-grained lexical cues that are typically averaged out in single-vector pooling.

\subsubsection{Hybrid Fusion Strategies}
One robust operational strategy for mitigating domain shifts and lexical blind spots is the deployment of hybrid architectures that ensemble dense and sparse signals. The integration of these modalities requires a rigorous merging framework to reconcile their divergent scoring distributions. BM25 scores are theoretically unbounded while dense cosine similarity is bounded within the range of $[-1, 1]$. Direct summation of these unnormalized values yields mathematically invalid rankings.

To address this, Reciprocal Rank Fusion (RRF) has emerged as the standard for orchestration. By discarding raw scalar scores in favor of rank positions, RRF functions as a robust non-parametric ensemble method. It effectively smoothes out the noise and failures of individual retrievers. For applications requiring more granular control, Linear Interpolation provides a parametric alternative. This method governs the contribution of each system via a weighting coefficient $\alpha$:

\begin{equation}
\text{Score}_{\text{hybrid}} = \alpha \cdot \text{Norm}(\text{Score}_{\text{dense}}) + (1 - \alpha) \cdot \text{Norm}(\text{Score}_{\text{sparse}})
\end{equation}

Research by \citet{shehata2022early} indicates that while an $\alpha \approx 0.5$ provides a strong general-purpose baseline, the optimal operating point is highly domain-dependent. Keyword-heavy retrieval tasks necessitate a lower $\alpha$ to prioritize sparse signals. In contrast, intent-heavy domains benefit from a higher $\alpha$ to leverage semantic understanding.

\section{Temporal Drift: The Silent Degradation of Retrieval Quality}
As established in the preceding analysis, modern dense retrievers are predominantly trained on static corpus snapshots. This effectively freezes their semantic worldview at the moment of model training. While this assumption of stationarity simplifies architectural design, it stands in direct contradiction to the open-ended and non-stationary nature of real-world information streams. This discrepancy precipitates temporal drift, which is a progressive divergence between the frozen model parameters and the evolving joint distribution of queries and documents. \citet{lazaridou2021mind} characterize this phenomenon as the temporal generalization gap. They observe that models show an increase in perplexity and retrieval degradation when evaluated on data emerging after the training period.

This section analyzes the mechanisms of this decay and the architectural paradigms devised to mitigate it. The challenge is framed as a fundamental tension in retrieval architecture: the trade-off between plasticity, which is the capacity to ingest new knowledge, and stability, which is the retention of established semantic alignments. Navigating this dichotomy requires strategies that prevent catastrophic forgetting without incurring the prohibitive costs of continuous model retraining.

\subsection{The Mechanics of Decay: Semantic and Distributional Shift}
Temporal drift constitutes more than the mere accumulation of novel documents; it represents a change in the joint distribution of queries and relevant documents. \citet{lazaridou2021mind} empirically characterize this as a temporal generalization gap. They demonstrate that standard evaluations inflate perceived robustness by neglecting the time-series nature of language data. When evaluated on corpora gathered post-training, the models show an increase in perplexity. This framework identifies two primary vectors through which this degradation propagates: Semantic Shift and Universe Shift.

\textbf{Semantic Shift.} Lexical semantics are inherently fluid, and term meanings evolve in response to cultural and sociopolitical events \cite{han-etal-2025-temporal, ishihara2022semantic}. \citet{ishihara2022semantic} quantify this volatility via Semantic Shift Stability. They observe that Large Language Models experience sudden performance declines during periods of high instability. For instance, tokens such as \textit{corona} or \textit{vaccine} underwent radical contextual realignments in 2020. Similarly, entities like \textit{Amazon} have historically drifted from natural to corporate semantics. These shifts render pre-trained embeddings obsolete because the geometric proximity learned during training no longer reflects current semantic equivalence.

\textbf{Universe Shift.} Distinct from lexical evolution is the shift in factual ground truth, which exposes the limitations of the standard bi-encoder architecture. In Dense Passage Retrieval (DPR), query and document encoders are optimized on a specific data snapshot. Consequently, the model weights $\theta$ encode the co-occurrence patterns of that era as static geometric truths. For example, a model trained on 2018 data learns to project the query "current US president" into the vector space region occupied by "Donald Trump." When the real-world state evolves, the frozen parameters $\theta$ fail to update the projection. This results in temporal misalignment where the model retrieves historically accurate but currently irrelevant documents. \citet{hinder2023model} provide a theoretical formalization of this failure mode by characterizing drift via drift localization. This involves the identification of specific regions in the data space where the conditional probability $P(Y|X)$ diverges. 

\subsection{Continual Learning and Incremental Adaptation}
To address the non-stationarity of data streams, researchers utilize Continual Learning (CL) paradigms. These enable models to ingest new knowledge without retraining from scratch. The central challenge in this domain is the plasticity-stability dilemma. This requires balancing the need to adapt to new tasks $T_{t}$ while preserving the weights necessary for $T_{t-1}$.

\subsubsection{Mitigating Catastrophic Forgetting}
The most direct adaptation strategy involves updating model parameters on incoming data. However, this introduces the risk of catastrophic forgetting, where optimization for the current distribution overwrites prior knowledge \cite{hou2025advancing}. \citet{hou2025advancing} formalize this within the Continual Learning Framework for Neural Information Retrieval (CLNIR). They evaluate the efficacy of regularization versus replay-based methods. While regularization techniques like Elastic Weight Consolidation (EWC) \cite{kirkpatrick2017overcoming} and Synaptic Intelligence (SI) \cite{zenke2017continual} offer theoretical constraints on weight updates, empirical findings suggest they struggle with the high dimensionality of embedding models. Instead, replay-based methods demonstrate superior robustness for pre-training based models. In this approach, specifically Naive Rehearsal (NR), the training set is a union of current data and a memory buffer of previous samples.

In scenarios where retaining raw data is infeasible due to privacy or storage constraints, distillation offers an alternative. Learning Without Forgetting (LwF) utilizes the previous model version as a Teacher and the updated version as a Student \cite{li2017learningforgetting}. By minimizing the divergence between the soft labels of the teacher and the outputs of the student on new data, LwF creates a temporal anchor. This forces the model to preserve its historical ranking logic while learning discriminative features for the new task.

\subsubsection{Curriculum Strategies and Temporal Bias}
The curriculum guiding the sequence of data presentation plays a critical role in shaping the temporal worldview of the retriever. \citet{santosh2024chronoslextimeawareincrementaltraining} investigate this in the ChronosLex framework. Their analysis reveals a trade-off between strict chronological ordering and random shuffling. Purely incremental training induces a Recency Bias, where the model adapts to emerging trends but degrades on historical precedents. Conversely, random shuffling obliterates temporal dependencies entirely. The optimal strategy identified is a Chronological Curriculum with Replay. This method processes data streams chronologically but injects a diverse reservoir of historical examples into current batches. This approach maintains the temporal breadth of the embedding space and ensures that novel concepts do not overwrite fundamental definitions.

\subsection{Temporal-Aware Architectures and Objectives}
While Continual Learning focuses on weight updates, a distinct architectural lineage seeks to render the model explicitly time-aware. These approaches critique the standard Language Model behavior of averaging conflicting facts, such as merging distinct presidencies into a single blurred representation \cite{dhingra2022time}.

\subsubsection{Explicit Temporal Modeling}
To resolve semantic collisions, researchers have introduced mechanisms to partition the knowledge base chronologically. \citet{dhingra2022time} propose Temporal Injection, which conditions the model on textual prefixes to improve calibration on future facts. \citet{abdallah2025tempretriever} advance this via fusion-based architectures in TempRetriever. This model moves beyond textual prompts to embed query and document timestamps directly into the dense representation space. This approach significantly outperforms standard baselines on time-sensitive benchmarks like ArchivalQA \cite{wang2022archivalqa} by treating time as a latent variable.

At a more granular level, \citet{rosin2022temporal} modify the Transformer mechanism through Temporal Attention. By conditioning self-attention scores on the time difference between tokens, the model dynamically weights context based on temporal proximity. This allows for semantic change detection without altering the input text structure. This enables the retriever to distinguish between the historical and modern contexts of an entity.

\subsubsection{Training Objectives and Geometric Compensation}
Adapting the retriever requires a re-evaluation of training objectives. \citet{abdallah2025tempretriever} argue for Temporal Hard Negatives, which are documents that were relevant in a past timeframe but are incorrect for the current temporal intent of the query. To enforce temporal reasoning during pre-training, \citet{han-etal-2025-temporal} introduce Temporal Span Masking (TSM). This targets time-specific information and forces the encoder to learn temporal dependencies.

Finally, the deployment of evolving models introduces a severe geometric constraint. If the model weights drift, the entire pre-computed index of billions of documents becomes invalid. \citet{goswami2025querydriftcompensationenabling} propose Query Drift Compensation (QDC) to resolve this without the prohibitive cost of re-indexing. QDC accepts embedding drift as inevitable and models it as a measurable transformation vector. This allows the system to geometrically align new queries with the existing older index space.

\section{Challenges and Future Directions}
\label{sec:future_directions}
Retrieval systems are at the center of many applications. As retrieval systems continue to evolve, we identify the following critical directions and areas for the evolution of retrieval architectures: 
\begin{itemize}
    \item \textbf{Coupled components:} Current retrieval stacks (chunking, re-ranking, query fan-out, etc) are highly interdependent. This forces components training and designs to consider the limitations and strength of other components. This poses a challenge during experimentation and iterations of the component, where other components in the pipeline could serve as bottleneck. Research to automatically identify such scenarios and detect them early could be beneficial for the real-world deployments involving large group of people working on different components simultaneously.
    
    \item \textbf{Adaptive Retrieval:} Static hyperparameters (e.g., fixed top-$k$, static reranking depth) result in computational inefficiency. Research should pivot toward dynamic orchestration, where the system estimates query complexity in real-time to allocate retrieval resources and pipeline depth adaptively. This is analogous to test-time scaling \cite{snell2024scaling, muennighoff2025s1}, which have shown promising improvements in generative applications.
    \item \textbf{Personalization:} Standard dense retrieval operates on a global relevance paradigm, treating the query as the sole context. However, relevance is often subjective and dependent on user history \cite{shah2025stateful, matthijs2011personalizing} and preferences \cite{shah2025user1}. While adding lightweight adapter layers or employing personalized re-ranking are potential solutions, the direct generation of personalized embeddings remains an underexplored but promising area and the impact of these logs as training data remains an interesting problem.
    
    \item \textbf{Generative Retrieval:} As query complexity exceeds the capacity of heuristic decomposition \cite{chatterji2025people, tamkin2024clio}, the paradigm may shift toward Generative Indexing (Differentiable Search Indices). Yet, this requires solving critical bottlenecks regarding incremental corpus updates and scaling to billion-parameter document spaces. Research and innovation in this direction would make these technique universal deployable on many use-cases involving non-static corpus.
    
    \item \textbf{Mechanistic Interpretability:} The opacity of dense vector spaces blocks debugging and headroom analyses. Developing methods to attribute retrieval scores to specific semantic features or training data is essential for ensuring fairness, trust, and auditability in deployment. In addition, the interpretability will also help during evaluation and identify opportunities and gaps.
\end{itemize}

\section{Conclusion}
This framework has systematically traversed the retrieval system design stack. It offers a unified perspective on the intricate trade-offs that define modern retrieval systems. The analysis began by deconstructing the Representation Layer, where the scalability of bi-encoders was contrasted with the deep semantic fidelity of Cross-encoders and hybrid architectures. The examination of the Granularity Layer demonstrated the necessity of thoughtful segmentation, ranging from atomic to hierarchical approaches, to mitigate information bottlenecks and preserve global context in long documents. 

In the Orchestration Layer, this review surveyed strategies that transcend single-vector limitations. It emphasized multi-view embeddings, query decomposition, and reasoning-driven hierarchical frameworks. This was followed by an analysis of multi-stage reranking pipelines that rely on computationally expensive but precise ranking strategies. Finally, the Robustness Layer highlighted the failure modes of embedding-based retrieval systems. This section detailed the need for architectural and hybrid approaches to combat challenges in domain generalization, exact-match failures, and the silent degradation caused by temporal drift. 

Building on this architectural breakdown, this framework identifies current research gaps and proposes forward-looking directions to inspire further study. The trajectory of retrieval within recommendation and generative ecosystems is poised for substantial evolution. It is expected that the integration of retrieval-augmented reasoning will shift from static matching to dynamic, agentic navigation of knowledge spaces. This work acts as a catalyst for that growth by fostering developments that refine retrieval capabilities and optimize the overall quality of user interactions.

\section*{Acknowledgments}

All study design, literature review, synthesis, and writing were conducted by the authors. Generative AI tools (Gemini) were used only for grammar checking, figures, and proofreading during the final polishing of the manuscript. No generative AI system was used to generate content, interpret prior work, or draw conclusions. The authors reviewed and approved all final text and remain fully responsible for the content of the paper.

\bibliographystyle{IEEEtranN}
\bibliography{retrieval_paper}

\begin{IEEEbiographynophoto}{Deep Shah}
is currently a Machine Learning Engineer at Google, leading key strategic initiatives in personalization for Google Search. He holds an M.S. degree in computer science from the University of Illinois Urbana-Champaign (UIUC). He has over nine years of industry experience specializing in retrieval systems, unsupervised learning, large-scale recommendation systems, and applications involving Large Language Models (LLMs). His work on Search products has received worldwide recognition for its impact and innovation.
\end{IEEEbiographynophoto}

\vskip 0pt plus -1fil

\begin{IEEEbiographynophoto}{Sanket Badhe} is a seasoned Machine Learning Engineer with over 10 years of experience specializing in AI Security, large-scale ML systems, and LLM applications. He currently leads key ML initiatives at Google for Youtube shopping. Sanket holds a Master’s in Data Science from Rutgers University and a B.Tech from IIT Roorkee, with prior experience at Tinder, TikTok, Oracle etc. His research has been published in ACL, CAMLIS etc.
\end{IEEEbiographynophoto}

\vskip 0pt plus -1fil

\begin{IEEEbiographynophoto}{Nehal Kathrotia} is currently a Software Engineer at Google, where she focuses on architecting secure, large-scale distributed systems for the Google Cloud Platform. She holds a B.Tech. degree in Computer Science from Dharmsinh Desai University. Her expertise spans Information Retrieval and Natural Language Processing, grounded in her research on Question Answering systems for the SQUAD dataset at IIT Bombay and her tenure with the Google Workspace Search team, where she led data sovereignty initiatives for search infrastructure. She has extensive industrial experience designing high-throughput systems capable of handling millions of queries per second.
\end{IEEEbiographynophoto}

\EOD
\end{document}